\documentclass[conference]{IEEEtran}
\IEEEoverridecommandlockouts
\usepackage{cite}
\usepackage{amsmath,amssymb,amsfonts}
\usepackage{algorithmic}
\usepackage{graphicx}
\usepackage{listings}
\usepackage{algorithm}
\usepackage{algorithmic}
\usepackage{textcomp}
\usepackage{xcolor}
\def\BibTeX{{\rm B\kern-.05em{\sc i\kern-.025em b}\kern-.08em
    T\kern-.1667em\lower.7ex\hbox{E}\kern-.125emX}}
\begin{document}

\newcommand{\qmcp}{\texttt{QMCPACK}}
\newcommand{\vecr}{\mathbf{r}}
\newcommand{\vecR}{\mathbf{R}}

\title{A High-Performance Design for Hierarchical Parallelism in the QMCPACK Monte Carlo code
\thanks{This manuscript has been authored in part by UT-Battelle, LLC, under contract
DE-AC05-00OR22725 with the US Department of Energy (DOE). The US government retains and the publisher, by accepting the
article for publication, acknowledges that the US government retains a nonexclusive, paid-up, irrevocable, worldwide
license to publish or reproduce the published form of this manuscript, or allow others to do so, for US government
purposes. DOE will provide public access to these results of federally sponsored research in accordance with the DOE
Public Access Plan (http://energy.gov/downloads/doe-public-access-plan).}
}

\author{\IEEEauthorblockN{1\textsuperscript{st} Ye Luo}
\IEEEauthorblockA{\textit{Computational Science Division} \\
\textit{Argonne National Laboratory}\\
Lemont, IL, USA \\
ORCID: 0000-0002-5117-2385}
\and
\IEEEauthorblockN{2\textsuperscript{nd} Peter Doak}
\IEEEauthorblockA{\textit{Computational Sciences and Engineering Division} \\
\textit{Oak Ridge National Laboratory}\\
Oak Ridge, TN, USA \\
ORCID: 0000-0001-6039-9752}
\and
\IEEEauthorblockN{3\textsuperscript{rd} Paul Kent}
\IEEEauthorblockA{\textit{Computational Sciences and Engineering Division} \\
\textit{Oak Ridge National Laboratory}\\
Oak Ridge, TN, USA \\
ORCID: 0000-0001-5539-4017}
}

\maketitle

\begin{abstract}
We introduce a new high-performance design for parallelism within the Quantum
Monte Carlo code QMCPACK. We demonstrate that the new design is better able to
exploit the hierarchical parallelism of heterogeneous architectures compared to
the previous GPU implementation. The new version is able to achieve higher GPU
occupancy via the new concept of crowds of Monte Carlo walkers, and by enabling
more host CPU threads to effectively offload to the GPU. The higher performance
is expected to be achieved independent of the underlying hardware, significantly
improving developer productivity and reducing code maintenance costs. Scientific
productivity is also improved with full support for fallback to CPU execution
when GPU implementations are not available or CPU execution is more optimal.
\end{abstract}


\section{Introduction}

QMCPACK, is a modern high-performance open-source Quantum Monte Carlo
(QMC)~\cite{becca_quantum_2017} simulation
code~\cite{kim_qmcpack_2018,kent_qmcpack_2020}. Its main applications are
electronic structure calculations of molecular, nanoscale and solid-state
systems. Variational Monte Carlo (VMC), diffusion Monte Carlo (DMC) and several
other advanced QMC methods are implemented with highly optimized algorithms.
These algorithms sample the positions of electrons within the simulated system
to accurately compute quantum mechanical properties. Their actual
implementations in QMCPACK are called QMC drivers. When QMCPACK development
started in the beginning of the first decade of this century, there were
initially only multi-threaded CPU drivers. With over 15 years of development,
QMCPACK has been well optimized to run on multicore CPU-only supercomputers like
Cori at the National Energy Research Scientific Computing Center (NERSC) and
Theta at the Argonne Leadership Computing Facility (ALCF). When NVIDIA GPUs
emerged in the field of high-performance computing, GPU dedicated QMC drivers
were introduced by using Compute Unified Device Architecture (CUDA). They
perform extremely well on supercomputers with a CPU-GPU hybrid architecture like
Summit at the Oak Ridge Leadership Computing Facility (OLCF), though at the cost
of portability. The CUDA-based GPU drivers are completely incompatible with
CPU-only drivers and call separate sets of subroutines of the wavefunction
calculation. This is required due to the different data layouts and algorithms
needed for high-performance on the GPUs\cite{esler2012}. For this reason, close
to redundant feature implementations are needed to satisfy both architectures.
If a feature implementation is missing in the GPU drivers, the whole simulation
needs to run with a CPU-only build because mixing CPU and GPU features is not
supported and the CPU/GPU selection is made at compile time.

As supercomputers start to reach Exascale and architectural diversity has
increased, and CUDA is no longer the only GPU programming model available. We
must find approaches to efficiently address the different
hardware\cite{ExascaleSkin2020}. The current Top500 rank-1 supercomputer, the
AMD-based Frontier at OLCF, generally prefers Heterogeneous-Computing Interface
for Portability (HIP) due to the installed AMD GPUs. To a large degree, CUDA
source code can be treated as HIP code directly, and be easily compiled for AMD
GPUs, making the portability issue not significant. However, on the Intel-based
Aurora machine at ALCF, SYCL is the preferred programming model for Intel
designed GPUs. The trick that treats CUDA as HIP won't work for SYCL considering
the fundamental difference between CUDA and SYCL. The QMCPACK developers have
already been struggling with maintaining both a CPU based and a CUDA based GPU
implementation. Adding additional drivers for each preferred programming model
is clearly not a sustainable direction.

In this work, we introduce a new universal design of batched QMC drivers which may
replace all the previous QMC drivers. The added flexibility in these drivers
enables maximizing code performance on specific hardware once users match
parallelism hierarchies properly to the actual software and hardware. The new
design supports the necessary data movement to allow mixing CPU-only and GPU
accelerated features to ensure a feature complete QMCPACK experience for the
user regardless of the hardware being used. Code specialization for specific
hardware remains possible for achieving potentially higher performance although
this no longer needs to be at the driver level.

This paper is organized as follows. Sec.~\ref{sec:drivers} analyzes the DMC
algorithm and its implementation before the new batched drivers are added. Sec.~\ref{sec:newbatcheddrivers}
introduces the details of the new drivers. Sec.~\ref{sec:results} shows how the code behaves in
CUDA-based GPU drivers and the new drivers, and discusses application performance.
Sec.~\ref{sec:conclusions} summarizes the hierarchical parallelism in QMCPACK.

\section{QMC drivers without the batched design}\label{sec:drivers}
\subsection{Basic QMC algorithm}
Before analyzing all the three sets of drivers, let us first understand
the characteristics of a DMC algorithm shown in Alg.~\ref{alg:DMC}.
\begin{algorithm}
  \begin{algorithmic}[1]
    \FOR{$\text{MC generation}=1\cdots M$}
      \FOR{$\text{walker}=1\cdots N_w$}
        \STATE let ${\bf R} = \{{\bf r}_1 \ldots {\bf r}_{N}\}$
        \FOR{$\text{particle}\  k=1 \cdots N$}
        \STATE set ${\bf r}_k^{\prime} \leftarrow {\bf r}_k+\nabla_k \Psi_T({\bf R)}+\delta$
           \STATE let ${\bf R}^{\prime} = \{{\bf r}_1 \ldots {\bf r}_k^{\prime} \ldots{\bf r}_{N}\}$
           \STATE \textbf{ratio} $\rho = \Psi_T ({\bf R}^{\prime})/\Psi_T ({\bf R})$
           \STATE \textbf{derivatives} $\nabla_k \Psi_T({\bf R}^{\prime})$
           \STATE Accept $\vecr_k \leftarrow \vecr_k^{\prime}$ or reject
        \ENDFOR \COMMENT{particle}
        \STATE \textbf{local energy} $E_L=\hat{H}\Psi_T({\bf R})/\Psi_T({\bf R})$
      \ENDFOR \COMMENT{walker}
      \STATE reweight and branch walkers based on $E_L - E_T$
      \STATE update $E_T$ and load balance via MPI.
    \ENDFOR \COMMENT{MC generation}
  \end{algorithmic}
\caption{Pseudocode for diffusion Monte Carlo.\label{alg:DMC}}
\end{algorithm}

\begin{itemize}
\item L1. The loop over generations is a sequential time-stepping loop for the
DMC imaginary time evolution.

\item L2. The walker evolution at each generation is independent of each other
and thus this loop can be parallelized. On parallel computers, walkers are first
parallelized over Message Passing Interface (MPI) and then parallelized within
each MPI process. Due to the fact that MPI is only needed for aggregating
results and handle walker count imbalance in L14, the parallel efficiency of QMC
algorithms over MPI can be made nearly perfect~\cite{kim_qmcpack_2018}, even at a scale of
thousands to millions of MPI processes. For the rest of this work, we restrict
the discussion of parallelization schemes of walkers within an MPI process.

\item L4. The loop of over particles (electrons) during random walking is also sequential.
Each iteration is called a single particle move since only one particle of a
walker is moved. This algorithm is referred to as particle-by-particle moves.

\item L5-6. Proposing a new electron position requires relatively cheap computation.
\item L7-8. When a single particle move gets proposed, heavy computational
  routines contain another vector loop over all the orbitals or particles.

\item L9. The computational cost depends on whether a proposed move gets
  accepted or rejected. Upon accepting a move, additional computation is needed
  to update the internal data of a walker, including a determinant matrix
  inverse which contributes the leading term of algorithmic complexity.
  Rejecting a move doesn't need computation and results near zero cost. This
  line causes the major computational cost difference at each single particle
  move. In VMC simulations, the acceptance ratio is typically between 20--80\%.
  However, within the more costly DMC, the acceptance ratio is usually very high ($>
  99\%$).
\item L11. Energy evaluations are required for every walker after single particle moves.  They are expensive.

\end{itemize}

\subsection{Multi-threaded CPU drivers}
\begin{algorithm}
  \begin{algorithmic}[1]
    \FOR{$\text{MC generation}=1\cdots M$}
      \STATE \#pragma omp parallel for
      \FOR{$\text{walker}=1\cdots N_w$}
        \FOR{$\text{particle}\  k=1 \cdots N$}
          \STATE ...
        \ENDFOR \COMMENT{particle}
      \ENDFOR \COMMENT{walker}
    \ENDFOR \COMMENT{MC generation}
  \end{algorithmic}
\caption{Pseudocode for the multi-threaded CPU implementation.\label{alg:CPU}}
\end{algorithm}

On multi-core CPUs, the multi-threaded CPU driver implementation distributes walkers
over CPU cores via OpenMP threads as shown in Alg.~\ref{alg:CPU}. The needed code change is minimal.
Although the cost of each single particle move depends on whether the proposed move
is accepted or rejected, the overall cost of each walker is almost equal once the single
particle move loop completes given the acceptance ratios across walkers.
Thus, the load-balance of threads is also near perfect. In the CPU implementation,
we also adopt OpenMP \texttt{simd} directives for the vector loop mentioned in Algo.~\ref{alg:DMC}
to leverage the Single instruction, multiple data (SIMD) units on modern CPUs~\cite{sc17}. This parallelization strategy works extremely
well on many-core wide vector CPUs including Intel Xeon, AMD EPYC and Fujitsu A64FX processors.
  
\subsection{CUDA-based GPU drivers}
The introduction of GPUs in HPC challenged the above parallelization strategy.
The accelerator characteristics of GPUs require sufficiently heavy compute kernels
to amortize kernel submission or synchronization cost in microseconds.
In the multi-threaded CPU drivers, each compute routine only handles the work of a single walker.
For large simulation problems ($>1000$ electrons), the workload of a single walker
can keep GPUs busy. But many QMCPACK users run small to medium problem sizes
($<=1000$ electrons) in scientific production, so dispatching GPU computation from
multi-threaded CPU drivers results in slow execution with most of the time being spent
in GPU overhead. For this reason, a GPU friendly scheme~\cite{esler2012} was
devised and implemented in the CUDA-based drivers in QMCPACK.

\begin{algorithm}
  \begin{algorithmic}[1]
    \FOR{$\text{MC generation}=1\cdots M$}
      \FOR{$\text{particle}\  k=1 \cdots N$}
         \STATE Algorithm 1. Line 5,6,7,8,9 over all the  $N_w$ walkers
      \ENDFOR \COMMENT{particle}
      \STATE \textbf{local energy} $E_L=\hat{H}\Psi_T({\bf R})/\Psi_T({\bf R})$ over $N_w$
      \STATE reweight and branch walkers based on $E_L - E_T$
      \STATE update $E_T$ and load balance via MPI.
    \ENDFOR \COMMENT{MC generation}
  \end{algorithmic}
\caption{Pseudocode for the CUDA-based implementation.\label{alg:GPU}}
\end{algorithm}

In Alg.~\ref{alg:GPU} it appears that the loop over walkers in
Alg.~\ref{alg:DMC} disappeared. Actually it is not removed but is added inside
each of the computational routines previously serving only one walker at a time.
All the computational routines on L3 now handle all the walkers in a batched
operation. As a result, all the walkers advance in lock-step for each single
particle move. All the compute kernels expose both vector computation and walker
concurrency and fit extremely well the hierarchical design of GPUs with threads
and thread blocks. For small simulated systems, GPUs have sufficient memory to
enable batching over hundreds to thousands of walkers. This is sufficient to
hide most of the GPU kernel overhead in practice.

However, this scheme has a few limitations:
(a) In the operation of accept/reject a single particle move, the number of walkers with
their proposed moves accepted must be large enough to avoid leaving
part of the compute hardware idle.
(b) There is only one thread enqueuing kernels and handling synchronization, while all the other
threads are idle. When the only working host thread is occupied
with handling of pre-/post-kernel processing, the GPU is also left idle.
(c) Most of the time, there is only one CUDA stream being used and there is limited overlap
between kernel execution and data transfer or concurrent kernel execution.
Using multiple CUDA streams can be added but requires significant code implementation.
(d) Assigning one thread block per walker only works for small problem sizes.
It doesn't allow leveraging more thread blocks per walker to further speed up the computation.
Meanwhile, large problems cannot keep many walkers resident on GPU
due to the device memory capacity limit and thus limits the full use of hardware resources.

\subsection{Deficiency of multi-threaded and CUDA-based drivers}
\begin{figure}
  \centering
  \includegraphics[width=0.95\columnwidth]{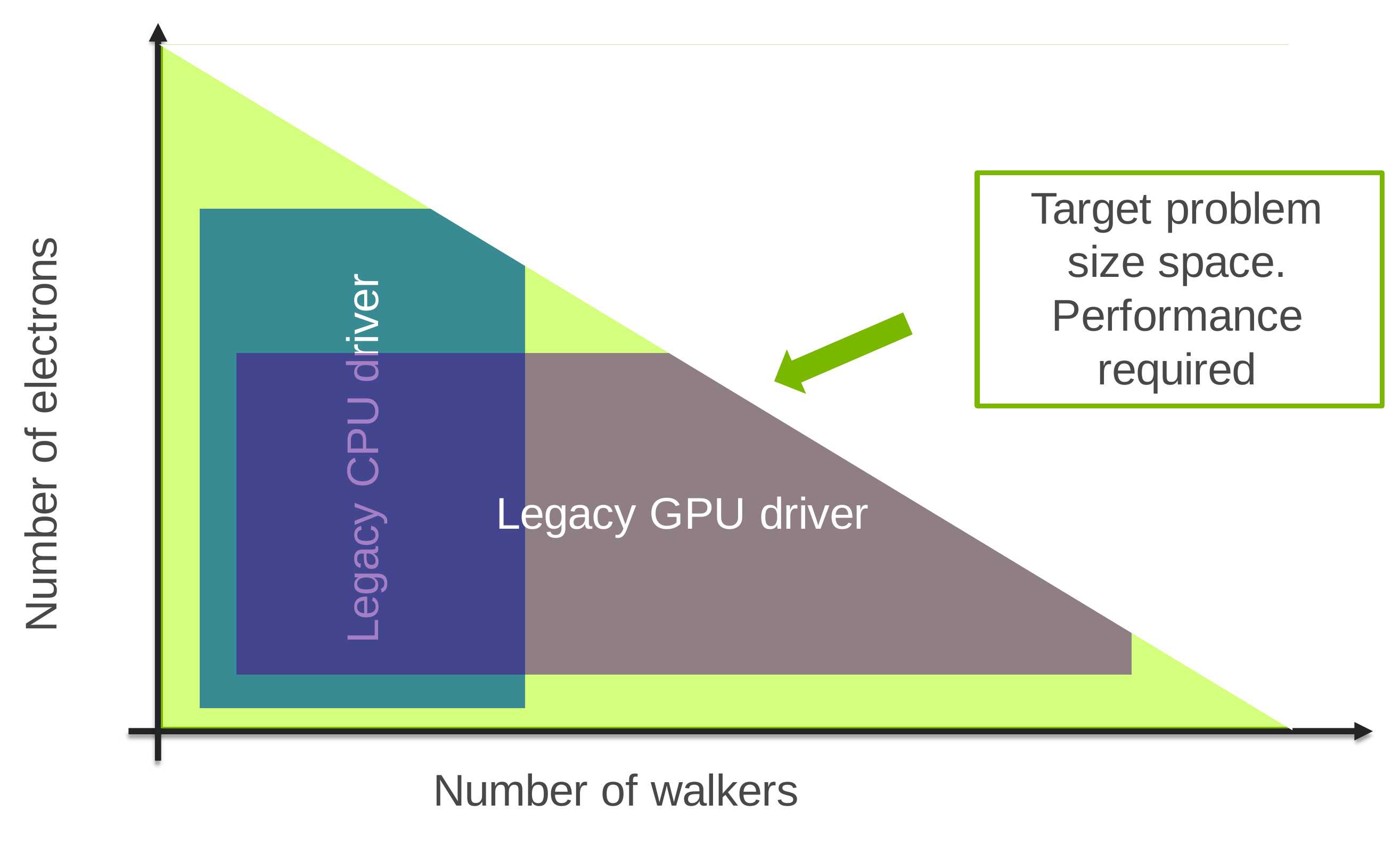}
  \caption{Parameter space of efficient runs with multi-threaded CPU (teal) or GPU (grey) driver.\label{fig:working_space_legacy}}
\end{figure}

Constrained by their pre-determined parallelization schemes, either of the above
driver designs only works efficiently in a limited parameter space as
illustrated in Fig.~\ref{fig:working_space_legacy}. The lower triangle parameter
space restriction comes from memory capacity limits. In addition, the
multi-threaded CPU drivers only invoke single walker APIs while CUDA-based
drivers only invoke ``batched'' or multi walker APIs. Due to the difference in
data layout and assumptions of data locations, it is not possible to fallback
from one implementation to the other and every feature must be implemented
separately for both multi-threaded CPU and GPU drivers. i.e. The CPU and GPU
codes were effectively internal forks of the codebase. This is clearly undesired due to the
additional developer effort and added maintenance cost. Only compute-heavy
features are worth porting to GPUs and running light computation on CPUs is
usually sufficient.

\begin{figure}
  \centering
  \includegraphics[width=0.95\columnwidth]{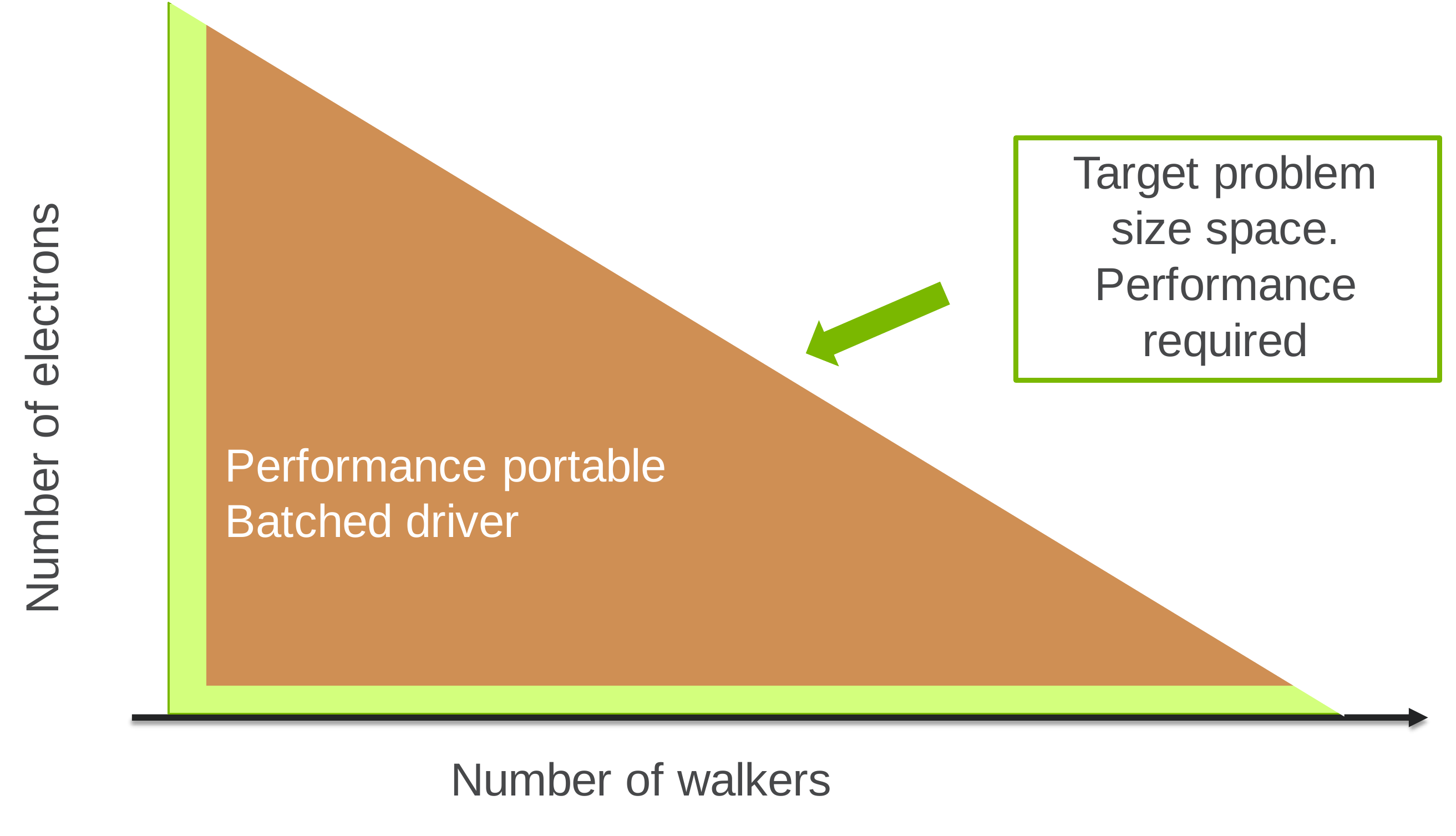}
  \caption{Parameter space of efficient runs with batched drivers (brown).\label{fig:working_space_pp}}
\end{figure}

With the above deficiency in mind, here we introduce a new high-performance
design for the QMC drivers and overall application. It avoids diverging code
paths at the driver level and works efficiently in the full possible parameter
space as illustrated in Fig.~\ref{fig:working_space_pp}.

\section{Performance portable batched drivers}\label{sec:newbatcheddrivers}
QMC drivers implemented in the new design are called batched drivers as walkers
are handled by compute devices in batches. This applies even for CPU based
machines, although a batch size of one is selectable, recovering close-to the
old CPU-only algorithm. Before explaining the design detail, we first introduce
a new concept ``crowd'', as a sub-organization of the walker population. A crowd
is a subset of the walkers that are operated on as a single batch. Walkers
within a crowd move through the operations of the QMC algorithm in lock-step. Walkers in different
crowds remain fully asynchronous unless operations over the full population
are needed.

The batched DMC driver pseudocode is shown in Alg.~\ref{alg:batch}. Compared to
the multi-threaded CPU implementation, the threaded loop over walkers has been
replaced with a threaded loop over crowds and thus crowds are fully parallelized
over host threads. The performance of multi-core CPUs can be easily maximized as
long as the number of crowds is chosen equal to the core count. When the crowd
size equal to 1, batched drivers behave exactly as multi-threaded CPU drivers.
With the crowd size larger than 1, the throughput of generating statistical
samples can potentially further increase due to improved data reuse or data
locality.

Compared to the CUDA-based implementation, the unchanged walker operation in
batches ensures dispatching sufficiently heavy computation in each GPU
invocation. The added multi-threaded crowds enable further improvements to GPU
utilization. When the number of crowd is restricted to one, batched drivers
behave exactly as CUDA-based drivers. With more than one crowd, crowds
parallelized over host threads concurrently sending operations to GPUs; data
transfer and kernel execution from different crowds may overlap if the
underlying hardware allows. Thus, the first three limitations of the CUDA-based
implementation are removed. Considering that the batched drivers do not mandate
specific data layouts, the last limitation of the CUDA-based implementation can
be removed by specializing compute kernels for extremely large problem sizes.

With the added crowds, batched drivers have a flexible number of batches and
batch sizes which can be tuned to maximize the performance of underlying
hardware. In the new driver design, the old set of walker batched computational
routines used by CUDA-based drivers are replaced with a new set which allow
falling back to computation using the single walker APIs. Consequently, batched drivers
allow mixing and matching CPU-only and GPU-accelerated features in a way that is
neither feasible with the multi-threaded CPU implementation nor the CUDA-based GPU one.

\begin{algorithm}
  \begin{algorithmic}[1]
    \FOR{$\text{MC generation}=1\cdots M$}
      \STATE \#pragma omp parallel for
      \FOR{$\text{crowd}=1\cdots C$}
        \FOR{$\text{particle}\  k=1 \cdots N$}
          \STATE Algorithm 1. Line 5,6,7,8,9 over all walkers with in this crowd
        \ENDFOR \COMMENT{particle}
        \STATE \textbf{local energy} $E_L=\hat{H}\Psi_T({\bf R})/\Psi_T({\bf R})$ over this crowd
        \STATE reweight and branch walkers based on $E_L - E_T$
        \STATE update $E_T$ and load balance via MPI.
      \ENDFOR \COMMENT{crowd}
CG    \ENDFOR \COMMENT{MC generation}
  \end{algorithmic}
  \caption{Pseudocode for the batched DMC driver.\label{alg:batch}}
\end{algorithm}

\section{Results}\label{sec:results}
\subsection{Demonstrating concurrent execution via GPU tracing}
\begin{figure}
  \centering
  \includegraphics[width=0.90\columnwidth]{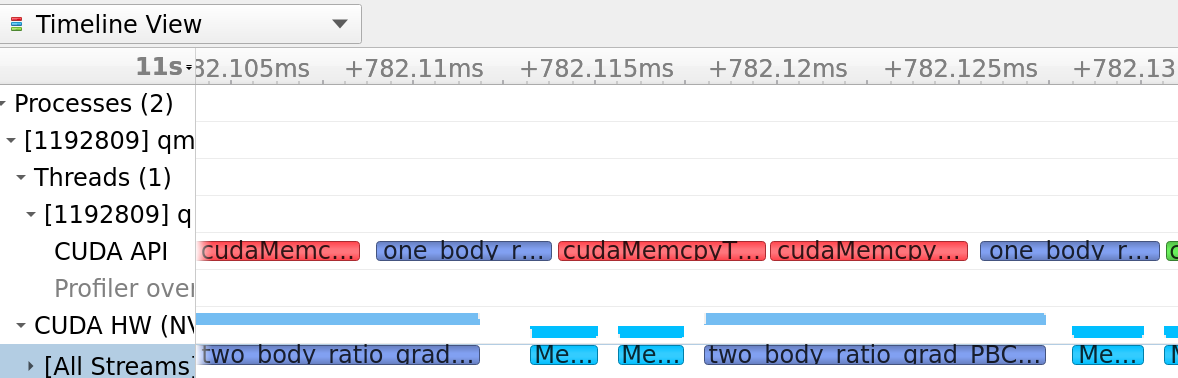}
  \caption{CUDA-based GPU driver GPU activity tracing. GPU API calls are made from a
  single thread and kernel execution and data
  transfers are all serialized.\label{fig:qmcpack_legacy_gpu}}
\end{figure}

\begin{figure}
  \centering
  \includegraphics[width=0.95\columnwidth]{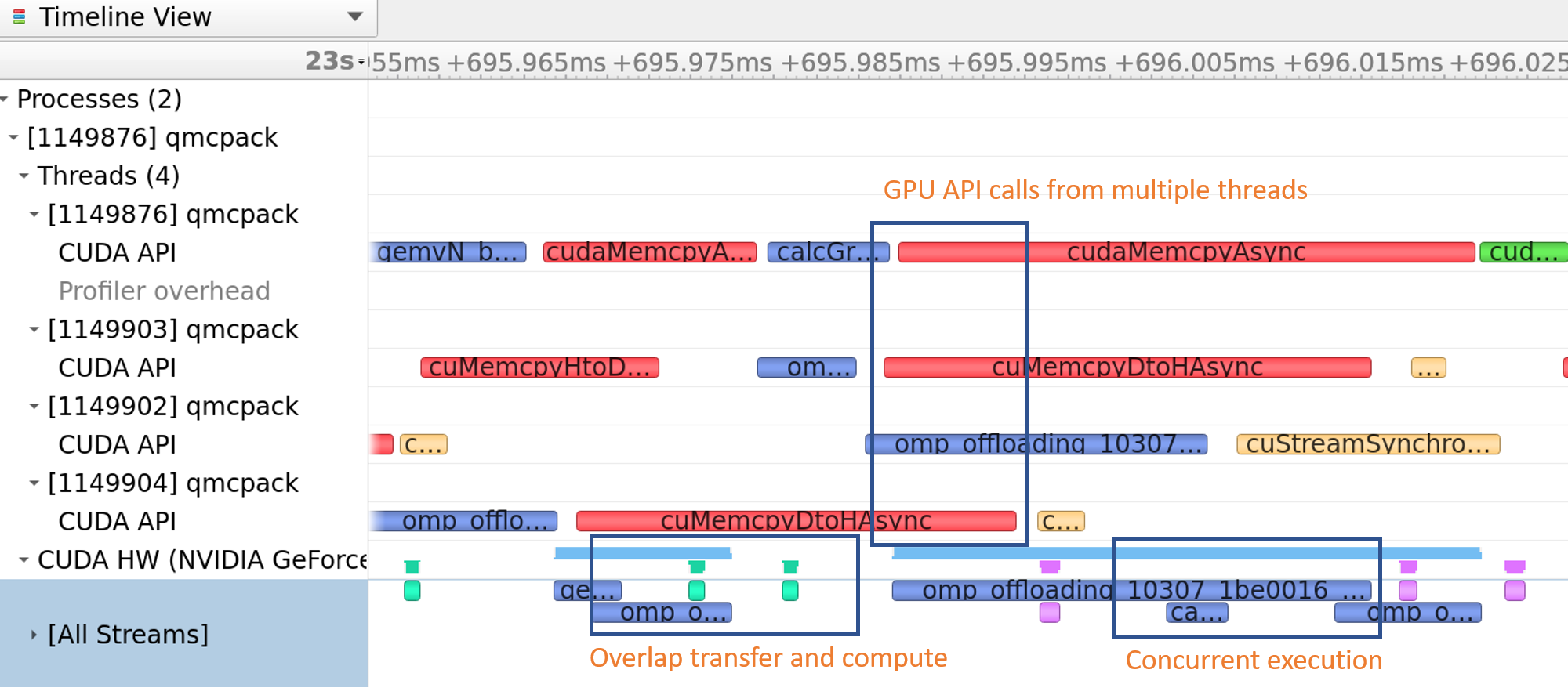}
  \caption{Batched driver GPU activity tracing. GPU API calls are made from multiple
  threads. Both concurrent kernel execution and overlapping kernel execution and data
  transfers are observed.\label{fig:qmcpack_batch_gpu}}
\end{figure}

In order to verify that batched drivers behave as expected on GPUs with real
simulations, we use NVIDIA Nsight Systems to trace GPU activities on an NVIDIA GPU
when running a QMCPACK performance test, which is a NiO 8-atom supercell DMC
simulation with 512 walkers. This is a ``small'' system where kernels are small
and therefore hiding kernel latency and maximizing concurrency is critical to
performance. Fig.~\ref{fig:qmcpack_legacy_gpu} shows the tracing of this test
using the CUDA-based DMC driver. There is only 1 thread active even though there
are 4 OpenMP threads available to the process. Kernel execution and data
transfer are serialized. Fig.~\ref{fig:qmcpack_batch_gpu} shows the tracing of
the same test using the new batched DMC driver. All the 4 OpenMP threads enqueue
kernels and submit data transfers to the GPU dedicated to this process. The GPU
keeps servicing requests from threads to maximize its utilization. Both
concurrent kernel execution and overlapping kernel execution and data transfers
are observed in the tracing. Higher efficiency is clearly obtained.

\subsection{Demonstrating gain in throughput on NVIDIA GPUs}
When targeting GPUs, the batched drivers currently use OpenMP offload and vendor
linear algebra libraries. Their performance are compared to the CUDA-based
drivers. The comparison is not apples to apples. In general, the batched drivers
have fewer pieces of features running on GPUs and additional data transfer can
be necessary between the GPU and the host. Although all the features used in the
NiO performance tests are accelerated in both the CUDA-based and batched
drivers, the implementations still differ due to the fundamental design change in the
batched drivers. And for the old CUDA-based drivers, all the kernels were
handwritten and highly optimized in CUDA while the performance of batched drivers
are affected by the quality of kernels generated by the OpenMP offload compilers.
Here we can compare the performance of both
drivers by the sampling throughput, namely the number of samples generated in a
given time. The study was conducted on the Summit supercomputer at Oak Ridge
National Laboratory. Each Summit node contains dual socket IBM Power 9
processors with 42 CPU cores in total and 6 NVIDIA V100 GPUs. The optimal way of
running QMCPACK requires 1 MPI rank per GPU. Thus, on each node, we place 6 MPI
ranks and each MPI process has its dedicated 7 CPU cores and 1 GPU.

\begin{figure}
  \centering
  \includegraphics[width=0.95\columnwidth]{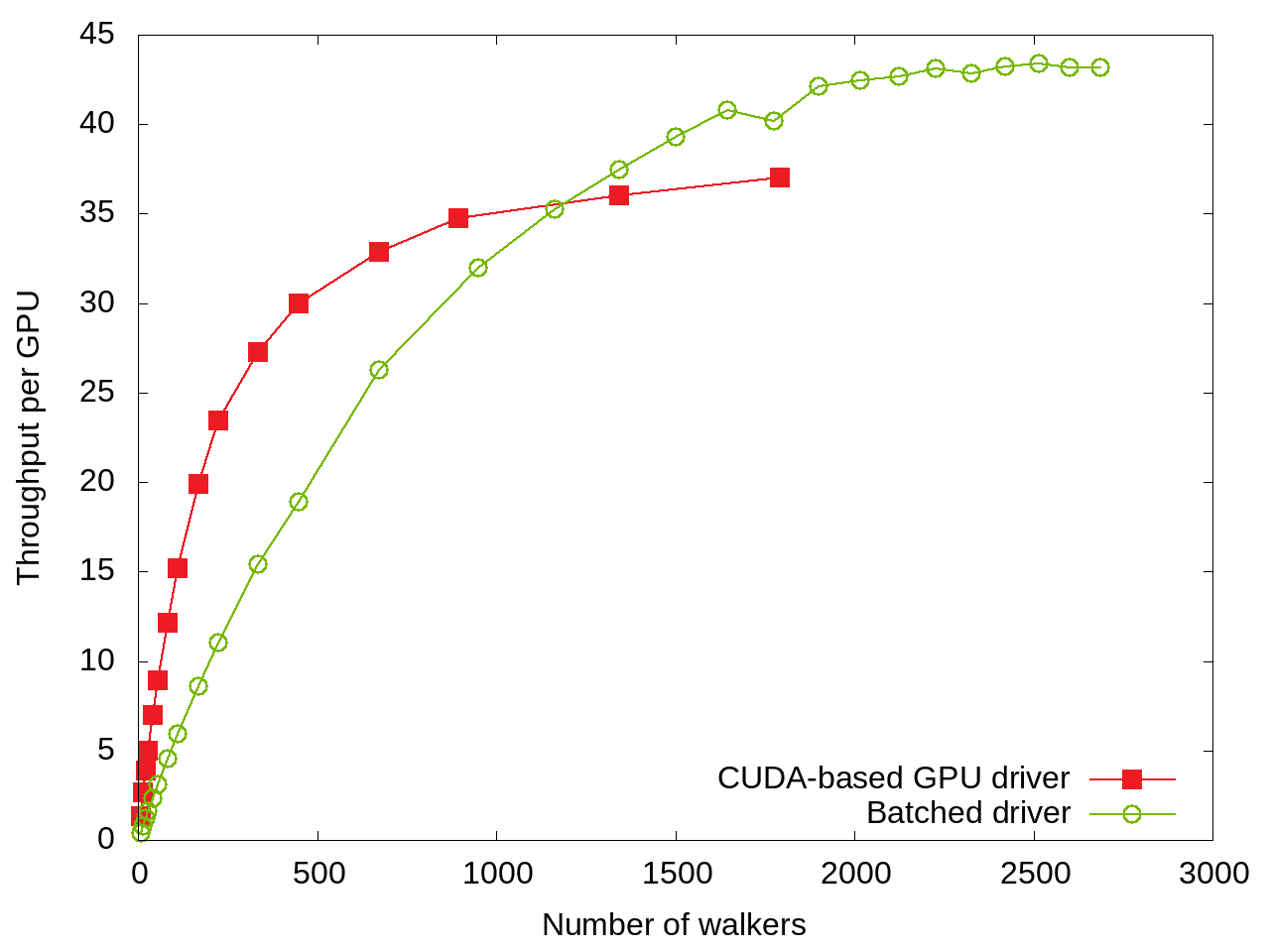}
  \caption{Sampling throughput as a function of walker counts in the 32 atom cell NiO solid DMC simulation.\label{fig:qmcpack_walker_scan}}
\end{figure}

Both the CUDA-based GPU drivers and batched drivers require optimizing the
walker count to maximize the throughput of a single GPU. Typically, the greatest
number of walkers prior to exhausting GPU memory is optimal. In
Fig.~\ref{fig:qmcpack_walker_scan}, the throughput of the runs with the
CUDA-based driver increases rapidly as walker count increases. It quickly
saturates at 1792 walkers once a single thread performance gets maxed out. With
the batched DMC driver, throughput grows slower in small walker counts. When the
total walker count per MPI rank is fixed, the walker batch size per thread is
smaller in batched drivers and thus in total more GPU overhead gets exercised by
all the threads. At larger walker counts when single thread maximal performance
is reached, batched drivers have more potential to maximize the full GPU
throughput by leveraging available threads. For this benchmark problem, the
measured performance becomes higher than the CUDA-based GPU driver when the
walker count exceeds 2000.

\begin{figure}
  \centering
  \includegraphics[width=0.95\columnwidth]{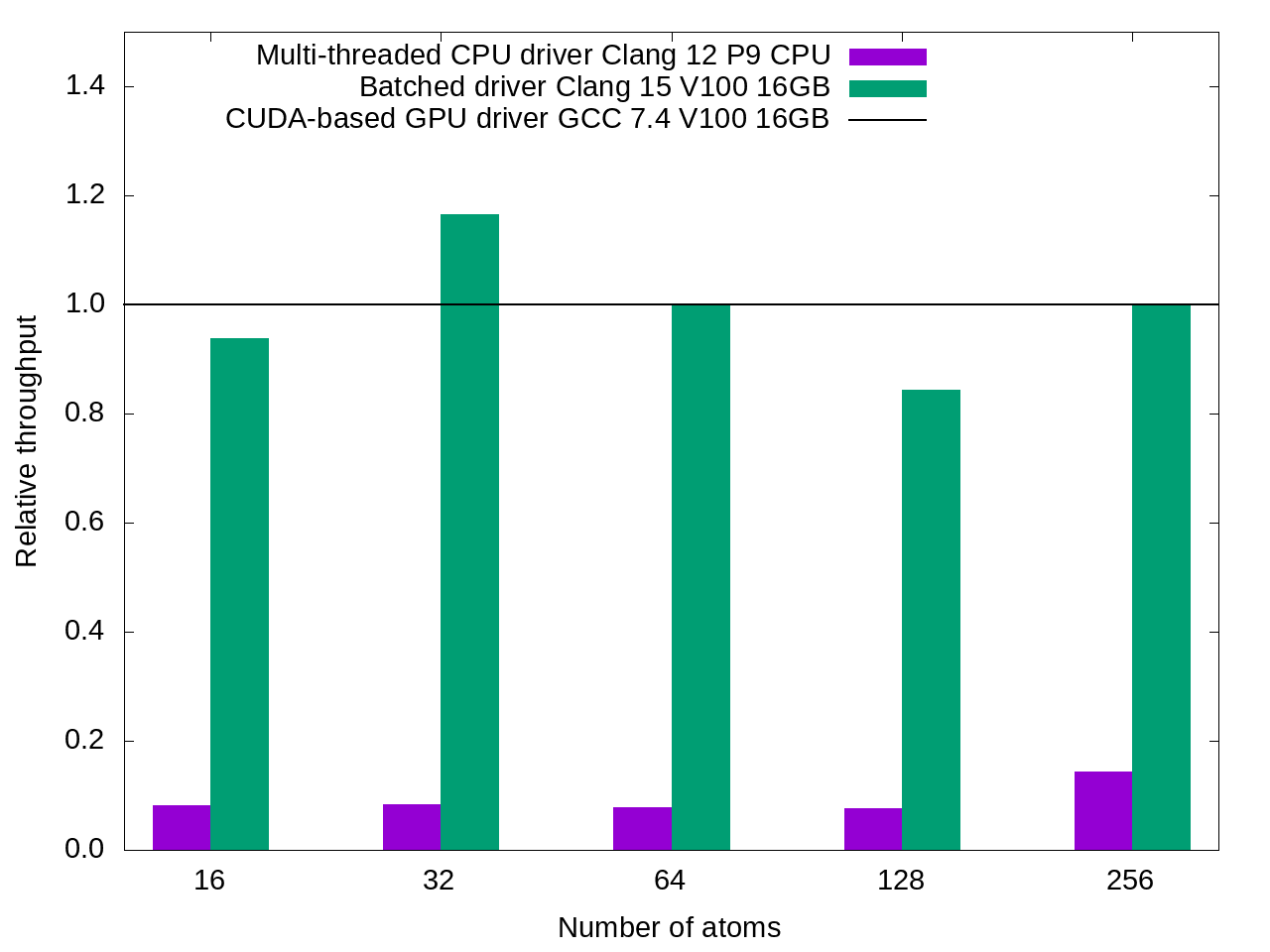}
  \caption{Batched driver throughput compared with the CUDA-based GPU driver.\label{fig:qmcpack_batch_gpu_throughput}}
\end{figure}

We also benchmark the code performance for a wide range of problem sizes.
The benchmark uses DMC simulations of NiO solids with 16, 32, 64, 128
and 256 atoms in the simulation cell. In Fig.~\ref{fig:qmcpack_batch_gpu_throughput} the throughput
of each problem size is rescaled by the throughput of CUDA-based GPU driver runs.
When running CPU-only, the relative throughput is only 10\% which reflects the fact that most of
compute power on Summit is from the NVIDIA V100 GPUs and the CUDA-based DMC driver is very well optimized.
Our newly designed batched DMC driver shows 80\% to 115\% relative performance depending on the problem size.
It is already suitable for scientific production simulations due to its competitive high-performance and feature complete nature.
With further optimization, we expect that the batched drivers will exceed the CUDA-based driver in all the benchmark cases.
As of August 2022, the compilers and libraries on non-NVIDIA platforms were not yet mature enough for benchmarking.
We plan to evaluate them once compilers and runtime libraries are sufficiently mature.

\section{Conclusion}\label{sec:conclusions}
Here we summarize the high-performance design of hierarchical parallelism in QMCPACK from the
coarse level to the fine level as implemented in the batched drivers.
\begin{enumerate}
\item Fully MPI distributed walker population. Usually one MPI per CPU socket or GPU.
  Extremely good strong and weak scaling across thousands to millions of compute nodes.
\item Multi-threaded crowds handle walkers within each MPI process.
  Each crowd does its independent time evolution. This is highly scalable on multi-core CPUs.
  In a CPU-GPU hybrid architecture, crowds may maximize the utilization of CPU cores
  before the shared GPUs are saturated by the workload.
\item Batched computation of walkers within each crowd.
  On GPUs, their computations can be submitted to GPUs with minimal GPU API overhead.
  On CPUs, there remains the possibility of breaking them into smaller tasks
  which can run on additional CPU threads if they are available. 
\item Compute kernels of each walker operate on a set of orbitals,
  usually the same or more than the electron count, or all the electrons.
  They can be fully vectorized on single instruction, multiple threads (SIMT)
  and single instruction, multiple data (SIMD) hardware. For extremely large problem sizes,
  these vector loop can be broken up in order to leverage more threads on
  CPUs or thread blocks on GPUs.
\end{enumerate}

By matching these parallelism levels to appropriate software abstractions in a
high-performance parallel computer, we believe that the maximal code performance
can be achieved regardless of the underlying hardware and true performance
portability can be achieved across CPU/GPU and even accelerators from any
vendor, potentially also including FPGAs and ASICs. In future work we plan to
demonstrate this portability.

\section*{Acknowledgment}
This research was supported by the Exascale Computing Project (17-SC-20-SC), a
collaborative effort of the U.S. Department of Energy Office of Science and the
National Nuclear Security Administration. This research used resources of the
Oak Ridge Leadership Computing Facility, which is a DOE Office of Science User
Facility supported under Contract DE-AC05-00OR22725. This research used
resources of the Argonne Leadership Computing Facility, which is a DOE Office of
Science User Facility supported under Contract DE-AC02-06CH11357.


\newpage
\bibliographystyle{IEEEtran}

\end{document}